\documentclass[lettersize,journal]{IEEEtran}
\usepackage{amsmath,amsfonts}
\usepackage{algorithmic}
\usepackage{algorithm}
\usepackage{array}
\usepackage[caption=false,font=normalsize,labelfont=sf,textfont=sf]{subfig}
\usepackage{textcomp}
\usepackage{stfloats}
\usepackage{url}
\usepackage{verbatim}
\usepackage{graphicx}
\usepackage{booktabs}
\usepackage{tabularx}
\usepackage{cite}
\usepackage{graphicx} 
\usepackage{subcaption} 

\begin{document}

\title{TTSlow: Slow Down Text-to-Speech with Efficiency Robustness Evaluations}

\author{Xiaoxue Gao,~\IEEEmembership{Member,~IEEE,}
        Yiming Chen,
        Xianghu Yue,~\IEEEmembership{Member,~IEEE,}
        Yu Tsao,~\IEEEmembership{Senior Member,~IEEE,}
        Nancy F. Chen,~\IEEEmembership{Senior Member,~IEEE,}
\thanks{Xiaoxue Gao and Nancy F. Chen are with Institute for Infocomm Research, Agency for Science, Technology, and Research (A*STAR), Singapore 138632 (e-mails: Gao\textunderscore Xiaoxue@i2r.a-star.edu.sg and nfychen@i2r.a-star.edu.sg).}
\thanks{Yiming Chen and Xianghu Yue are with the Department of Electrical and Computer Engineering, National University of Singapore, National University of Singapore, Singapore 117583 (e-mails: yiming.chen@u.nus.edu and xianghu.yue@u.nus.edu).}
\thanks{Yu Tsao is with Research Center for Information Technology Innovation, Academia Sinica, Taipei 115, Taiwan (e-mail: yu.tsao@citi.sinica.edu.tw).}
}
\maketitle
\begin{abstract}
Text-to-speech (TTS) has been extensively studied for generating high-quality speech with textual inputs, playing a crucial role in various real-time applications.
For real-world deployment, ensuring stable and timely generation in TTS models against minor input perturbations is of paramount importance. 
Therefore, evaluating the robustness of TTS models against such perturbations, commonly known as adversarial attacks, is highly desirable. 
In this paper, we propose TTSlow, a novel adversarial approach specifically tailored to slow down the speech generation process in TTS systems.
To induce long TTS waiting time, we design novel efficiency-oriented adversarial loss to encourage endless generation process.
TTSlow encompasses two attack strategies targeting both text inputs and speaker embedding. 
Specifically, we propose TTSlow-text, which utilizes a combination of homoglyphs-based and swap-based perturbations, along with TTSlow-spk, which employs a gradient optimization attack approach for speaker embedding. 
TTSlow serves as the first attack approach targeting a wide range of TTS models, including autoregressive and non-autoregressive TTS ones, thereby advancing exploration in audio security. 
Extensive experiments are conducted to evaluate the inference efficiency of TTS models, and in-depth analysis of generated speech intelligibility is performed using Gemini. 
The results demonstrate that TTSlow can effectively slow down two TTS models across three publicly available datasets. 
We are committed to releasing the source code upon acceptance, facilitating further research and benchmarking in this domain.
\end{abstract}


\section{Introduction}
\IEEEPARstart{T}{ext}-to-speech (TTS) is a task that aims to generate spoken speech based on a given text input \cite{yasuda2023text,chen2023vector,khanam2022text,liu2024text}. 
TTS has achieved remarkable advancements from the autoregressive TTS models \cite{wang2017tacotron,shen2018natural,ping2018clarinet,ping2018deep,cooper2020zero} to the non-autoregressive TTS approaches \cite{kim2020glow,ren2019fastspeech,ren2020fastspeech,kim2022guided,pratap2023scaling,kim2021conditional,casanova2022yourtts,ju2022trinitts,miao2020flow,peng2020non} and neural vocoders \cite{vanwavenet,prenger2019waveglow,kalchbrenner2018efficient,kumar2019melgan,yamamoto2020parallel} as well as the integration of diffusion models \cite{popov2021grad,huang2022prodiff,huang2023prosody,li2024styletts,li2023diclet} in recent years. 
These developments have led to diverse applications in virtual assistants, audiobooks, voice-over narration, and navigation systems \cite{li2023styletts,arik2017deep,luo2019emotional,sigurgeirsson2024controllable}.

Real-time inference capability is crucial for a TTS system to be deemed production-quality \cite{skerry2018towards,vasquez2019melnet}, as without it, the system becomes impractical for most TTS applications~\cite{arik2017deep}. Efficiency robustness and audio security are critical components of various practical applications of TTS systems \cite{haque2023slothspeech,petracca2015audroid}.
Notably, excessively long speech generation in TTS systems can even significantly compromise security in various domains. 
In banking, delayed speech synthesis may hinder transaction verification and security notifications, which give fraudsters more time to exploit vulnerabilities, posing a heightened risk to bank security and customer assets~\cite{melin2001ctt,yu2021methods}.
In home security, slow TTS can delay critical alerts (e.g., intruder detection, fire alarms), reducing system effectiveness and increasing the risk of harm and property damage~\cite{petracca2015audroid}. 
In car security, prolonged speech generation can delay navigation instructions and critical alerts, posing safety hazards~\cite{bisio2018smart,liu2017vehicle}.
Thus, TTS models must prioritize high-efficiency robustness to ensure usability, practicality, and effective security measures.
However, current efficient TTS works mainly focus on improving inference speed with normal inputs~\cite{ren2019fastspeech,ren2020fastspeech,peng2020non}, while the robustness of TTS models against minor perturbations remains largely unexplored.

Adversarial attack serves as a common and effective practice to evaluate the robustness of neural models recently for real-world applications~\cite{haque2023slothspeech,petracca2015audroid}.
Adversarial attacks aim to elicit incorrect predictions through slightly altering the input data, thereby enabling the automatic detection of the flaws in existing neural models and revealing their vulnerabilities~\cite{chen-etal-2023-dynamic,li2023white,ebrahimi2018hotflip,li2019textbugger}. This attack process, in turn, aids in assessing and enhances neural model robustness \cite{li2023sibling,chen2022nicgslowdown,haque2023slothspeech,olivier2022recent}. 
Motivated by the success of adversarial attacks and the under-studied robustness issue of TTS models, we propose TTSlow, a simple yet effective unified adversarial approach, to examine whether existing TTS models can provide stable and timely responses when confronted with maliciously perturbed inputs, termed as efficiency robustness.

Based on the observations that longer speech outputs lead to more inference steps and excessively long generation time in TTS models, we design TTSlow to automatically discover malicious inputs that can elicit endless speech generation process through nearly imperceptible input perturbations.
Specifically, TTSlow consists of two novel attack techniques: TTSlow-text, which incorporates both character-swap attack and homoglyphs replacement attack, and TTSlow-spk, achieved through a speaker-oriented projected gradient descent attack.
Extensive experiments on both autoregressive and non-autoregressive TTS models on three datasets show that both TTSlow-text and TTSlow-spk can significantly harm the inference efficiency with longer speech, longer inference time, higher inference energy and an average attack success rate (ASR) of 90.39\%.

The contributions of this paper include:
\begin{itemize}
    \item \textbf{New Problem Characterization}: We identify a novel audio security problem within the TTS domain, focusing on the automatic detection of model flaws to advance the field towards more trustworthy TTS systems. To the best of our knowledge, this is the first study to systematically investigate into the robustness of inference efficiency in TTS systems.
    \item \textbf{Novel approaches}: We propose novel optimization-oriented loss objectives for both text and speaker embedding-based adversarial attacks. Our TTSlow approach is also designed to encompass both autoregressive and non-autoregressive scenarios in TTS models, offering distinct objective designs tailored to each scenario. This work represents the first successful implementation of near-human imperceptible adversarial attacks on TTS systems.
    \item \textbf{Comprehensive Experimentation}: We conduct a systematic evaluation of two proposed attack strategies on two TTS models across three publicly available datasets. Our findings underscore the need for future research aimed at enhancing and safeguarding the inference efficiency robustness of TTS models.
\end{itemize}

The rest of this paper is organized as follows. Section~\ref{Related Work} describes related work on TTS and adversarial attacks, providing context for the proposed model design. Section~\ref{TTSlowSection} presents an overview of our proposed TTSlow approach and two proposed attack strategies. Section~\ref{TTSlow Objective Function} introduces the proposed objective functions for TTSlow in both autoregressive and non-autoregressive scenarios. Section~\ref{Experiments} details the database and experimental setup. Section~\ref{Results and Discussion} discusses the experiment results. Finally, Section~\ref{Conclusion} concludes the study.

\section{Related Work}
\label{Related Work}

We review text-to-speech models and adversarial attacks to establish the foundation of this work.
\subsection{Text-to-speech Models}

Text-to-speech (TTS) has garnered significant attention in recent times, powering numerous real-world applications \cite{ren2019almost,miao2021efficienttts,tan2024naturalspeech,guan2024reflow}.
In our investigation of the factors affecting TTS models' inference speed in practical applications, we delve into the length prediction mechanisms during the speech generation process.
The determination of when a TTS model should stop generating speech distinguishes TTS models into two primary categories: autoregressive and non-autoregressive models \cite{kim2020glow,ren2019fastspeech,ren2020fastspeech}.
Autoregressive TTS models typically stop speech generation based on the prediction of a stop token \cite{wang2017tacotron,shen2018natural,ping2018clarinet,ping2018deep,cooper2020zero} , whereas non-autoregressive TTS models determine the length of the generated speech using a duration predictor \cite{kim2020glow,ren2019fastspeech,ren2020fastspeech,kim2022guided,pratap2023scaling,kim2021conditional,casanova2022yourtts,ju2022trinitts}. 

Among these models, SpeechT5 stands out as a powerful autoregressive model, structured on an encoder-decoder architecture \cite{ao2022speecht5}. It demonstrates exceptional speech quality through extensive pre-training on large-scale unlabeled speech and text data, showcasing its potential for real-world applications. Conversely, VITS has gained fame and widespread adoption as a non-autoregressive model, utilizing variational inference with adversarial learning in an end-to-end TTS system \cite{kim2021conditional,pratap2023scaling}. These models, SpeechT5 and VITS, are selected as representative backbone TTS models for this study due to their distinct approaches and strong performance in their respective categories.

\begin{figure*}[t]
\centering
\includegraphics[width=179mm]{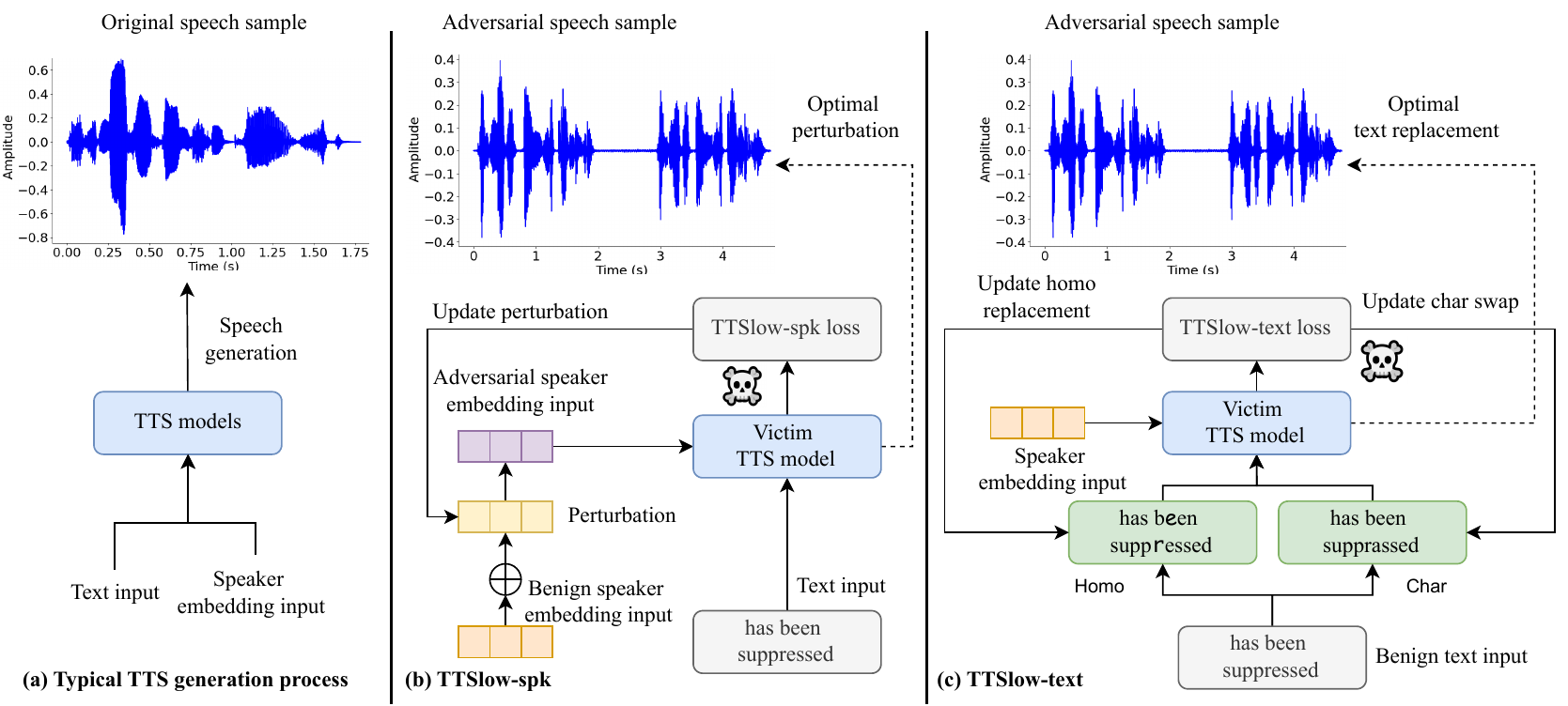}
\caption{The overview network architecture of (a) typical TTS generation process and the proposed TTSlow with two attack approaches: (b) TTSlow-spk and (c) TTSlow-text.}
\label{proposeFig}
\end{figure*}

\subsection{Adversarial Attacks}
Recently, the adversarial attack methods has been developed to evaluate the efficiency robustness of the machine-learning models for real-world applications, and it has been intensively studied in computer vision \cite{li2023sibling,chen2022nicgslowdown}, machine translation \cite{chen2022nmtsloth}, natural language processing~\cite{chen-etal-2023-dynamic,li2023white,ebrahimi2018hotflip,li2019textbugger}, automatic speech recognition \cite{haque2023slothspeech,olivier2022recent,schonherr2018adversarial,wang2022query,zhang2022investigating,ge2023advddos} and speaker identification \cite{zuo2024advtts,lan2022adversarial} domains.

Adversarial attacks can be classified into accuracy-oriented and efficiency-oriented attacks~\cite{li2023sibling,chen2022nicgslowdown,chen2022nmtsloth,chen-etal-2023-dynamic,li2023white,ebrahimi2018hotflip,li2019textbugger}. 
Accuracy-based attacks target reducing the robustness of models by decreasing recognition performance, as seen in adversarial attacks on speech recognition models~\cite{haque2023slothspeech,olivier2022recent,schonherr2018adversarial,wang2022query,zhang2022investigating,ge2023advddos}. 
Efficiency-oriented attacks, like SlothSpeech~\cite{haque2023slothspeech}, focus on diminishing system inference efficiency, significantly impacting speech recognition model performance.
However, there is a notable lack of research exploring the resilience of TTS models against security threats (adversarial attacks) from both accuracy and efficiency perspective. 
This paper aims to bridge the existing gap by proposing an adversarial attack method to analyze efficiency robustness for TTS systems.

\section{TTSlow}
\label{TTSlowSection}

In this section, we formulate the research problem and describe our proposed TTSlow approach, which includes two attack strategies: TTSlow-spk and TTSlow-text.

\subsection{Problem Formulation}
TTSlow aims to accomplish two objectives: (i) significantly increasing the computational time and reducing the inference efficiency for the victim TTS model, and (ii) maintaining minimal perturbations in the generated output. With these goals in mind, we approach the problem as one of constrained optimization problem:
\begin{equation}
    \label{eq:problem}
        \Delta = \mathop{argmax}\limits_{\delta}len_{f}(x+\delta) 
        \quad  s.t. ||\delta|| \leq \epsilon, 
\end{equation}
where $x$ represents the given benign input, and $\epsilon$ denotes the maximum allowed adversarial perturbation. $f$ is the victim TTS model and $len_{f}(\cdot)$ is the output sequence length of the victim TTS model. Our proposed approach, TTSlow, seeks to find the optimal perturbation $\Delta$ that reduces the efficiency while ensuring that the perturbation remains within the permissible threshold $\epsilon$ (i.e., nearly human unnoticeable).

\subsection{Overview of TTSlow}
\label{overview}
Fig.~\ref{proposeFig} shows an overview of proposed TTSlow against a typical multi-speaker TTS model.
Conditioned on both benign text input and speaker embedding, the trained TTS models successfully synthesize approximately 1.75 seconds of synthesized speech, as depicted in Fig.~\ref{proposeFig} (a).
Yet, when small perturbation is added to either the benign speaker embedding (TTSlow-spk in Fig.~\ref{proposeFig} (b)) or the benign text input (TTSlow-text in Fig.~\ref{proposeFig} (c)),
the generated speech length suddenly increases to approximately 4.5 seconds, leading to longer generation process and a decrease in inference efficiency.
The TTSlow approach employs two attack techniques to modify the given inputs and craft adversarial examples in Fig~\ref{proposeFig}. 
The proposed attack techniques are tailored for various TTS scenarios: TTSlow-spk suits multi-speaker models with speaker embedding, TTSlow-text is applicable to both multi-speaker and single-speaker models using characters as inputs. 
These attack techniques will be further detailed in the subsequent sections.



\subsection{TTSlow-spk}
\label{TTSlow-spk}
We begin by introducing the TTSlow-spk to assess the inference efficiency robustness of victim TTS models, as depicted in Fig.~\ref{proposeFig} (b).
TTSlow-spk aims to generate an adversarial speaker embedding sample \( s_{\text{adv}} \) using projected gradient descent \cite{madry2017towards,olivier2022there} as the optimization approach while keeping the input text fixed.

\subsubsection{Gradient Optimization}
\label{Gradient Optimization}
To slow down the speech generation, we propose to make the victim TTS model generate longer adversarial speech through updating the perturbation and optimizing the TTSlow-spk loss, denoted as $\mathcal{L}_{\text{TTSlow-spk}}$, aiming for no-stop speech generation. The TTSlow-spk attack is executed through iterating perturbations using gradient optimization with respect to the loss function $\mathcal{L}_{\text{TTSlow-spk}}$. 
The updated perturbation $\delta$ for each iteration is computed as follows:
\begin{equation}
\centering
\begin{split}
s_{adv}=s + \delta,\\
\delta \leftarrow \Pi \left\{ \delta -\alpha \cdot sign(\bigtriangledown_{\delta} \mathcal{L}_{\text{TTSlow-spk}}(f(s + \delta))) \right\}
\end{split}
\label{ttslow-spk loss}
\end{equation}
where s denotes the input speaker embedding, and \( \alpha \) is the learning rate. \( \bigtriangledown_{\delta} \mathcal{L}_{\text{TTSlow-spk}} \) is the gradient of the TTSlow-spk loss with respect to the perturbation \( \delta \). \( \Pi \{\cdot\} \) is the projection function that enforces the \( \ell_p \) constraint on the perturbation:
\begin{equation}
\Pi_{\ell_p}(s_{\text{adv}}) = \arg\min_{z \in S} \| s_{\text{adv}} - z \|_p,
\end{equation}
where \( z \) is an element within the feasible set \( S \), and \( \| \cdot \|_p \) denotes the \( \ell_p \) norm. \( \Pi \{\cdot\} \) maps \( s_{\text{adv}} \) to the closest point \( z \) in \( S \) under the \( \ell_p \) norm. We consider both \( \ell_{2} \)  and \( \ell_{inf} \) for distance norm. 

\subsubsection{Attack Methodology}
After explaining the gradient optimization attack mathematically, Fig.~\ref{proposeFig} (b) visually depicts the process of adding perturbation to the benign speaker embedding input, leading to the creation of adversarial speaker embedding \( s_{\text{adv}} \).
This adversarial embedding is then fed into the victim TTS model to compute the TTSlow-spk loss. Further details regarding the TTSlow-spk loss will be provided in Section \ref{TTSlow Objective Function}. 
After multiple iterations, the TTSlow-spk seeks to identify the optimal perturbation $\delta$ that minimizes the loss $\mathcal{L}_\text{TTSlow-spk}$ (thereby reducing the efficiency) while adhering to the perturbation constraint to remain the allowed threshold. This optimization process also aims to achieve near imperceptibility to humans, as measured by the distance norm.

\subsection{TTSlow-text}
\label{ttslow-text}
Inspired by novel text replacement approaches in NLP~\cite{ebrahimi2018hotflip,boucher2022bad}, we propose TTSlow-text, as illustrated in Fig.~\ref{proposeFig} (c), which combines two text-oriented attacks, homoglyphs replacement attack, and character swap attack. Homoglyphs are unique characters that render the same glyph or a visually similar glyph. TTSlow-text is designed to maintain the original text length while generating adversarial samples that are almost imperceptible to humans.

\subsubsection{Attack Methodology}
Text-oriented attacks specifically target text inputs while keeping the speaker embedding input unchanged. TTSlow-text is proposed to iteratively modify the given text inputs to create adversarial examples in two strategies towards the TTSlow-text loss $\mathcal{L}_{\text{TTSlow-text}}$, as depicted in Fig.~\ref{proposeFig} (c).  These two strategies include the character swap (char) attack, which replaces one character in the input text with another randomly selected character from the original character vocabulary, and the homoglyphs replacement (homo) attack, which replaces a character with its corresponding homoglyph. The homoglyph character mapping used in TTSlow-text follows the default mapping from TextBugger \cite{li2019textbugger}. The number of characters for replacement is decided by a fixed ratio of the input character length, and we set the ratio to 0.05.

\subsubsection{Differentiable Objective Approximation}
\label{Differentiable Objective Approximation}
Unlike attacking input speaker embedding, TTSlow-text encounters a non-differentiable issue since the input text is not differentiable in our optimization objective, as shown in Equation \ref{eq:problem}.
Therefore, we propose to design a differentiable objective to approximate our adversarial goals. 

We consider replacing the original character with another character $\hat{t}$ to achieve the optimal perturbation $\delta$:
\begin{equation}
        \delta = \mathop{argmax}_{\hat{t}} \ \text{Inc}_{t, \hat{t}},
\end{equation}
To compute the target character, we define character replace increment $Inc_{t,\hat{t}}$ to measure the efficiency degradation caused by replacing character $t$ to $\hat{t}$:
\begin{equation}
\label{ttslow-text loss}
        \text{Inc}_{t, \hat{t}} = \sum_{j} (E(\hat{t})  - E(t))_{j} \times \frac{\partial \mathcal{L}_\text{TTSlow-text}(t)}{ \partial t_i^j},
\end{equation}
where t denotes the input text and $E(\cdot)$ represents the text embedding vector of a given token. We note that $E(\cdot)$ is differentiable. $\text{Inc}_{t,\hat{t}}$ denotes the increase in the gradient of our objective function, resulting from replacing token $t$ with token $\hat{t}$. $\mathcal{L}_{\text{TTSlow-text}}$ is to encourage TTS model to generate non-stop speech frames, thereby slowing down the speech synthesis process. Further details regarding the TTSlow-text loss will be provided in Section \ref{TTSlow Objective Function}. 

\begin{figure}[t]
\centering
\includegraphics[width=97mm]{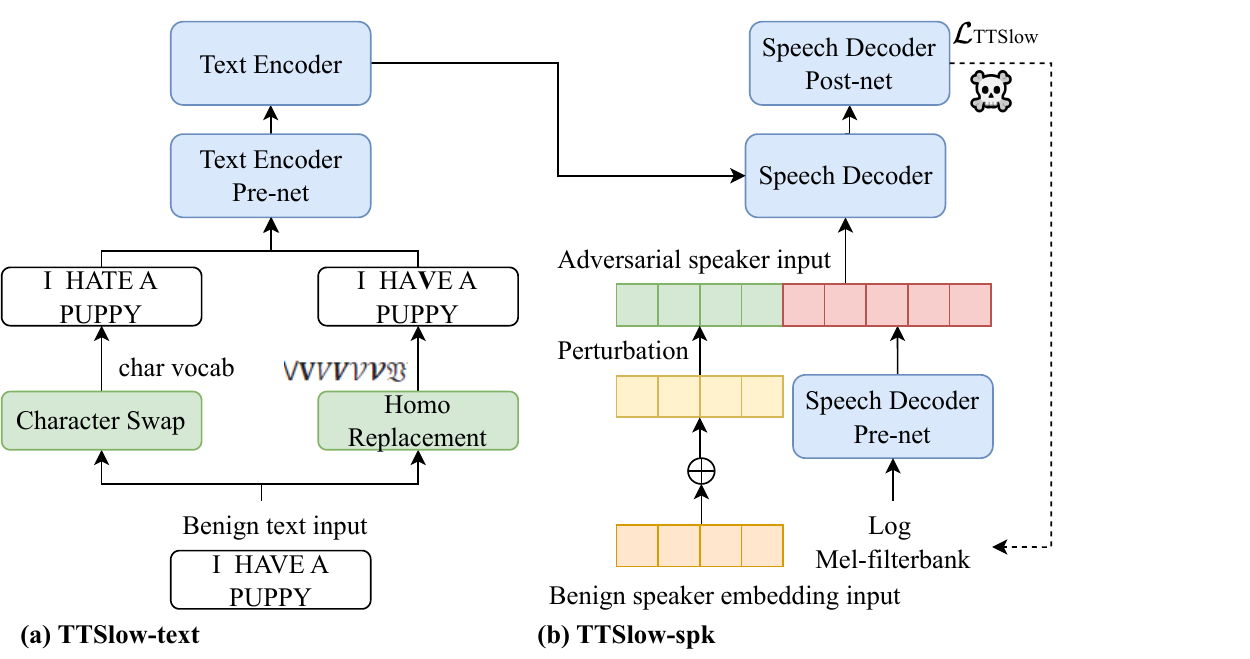}
\caption{Overview of TTSlow with TTSlow-text and TTSlow-spk Attacks on the autoregressive SpeechT5-tts Model}
\label{speecht5}
\end{figure}

\subsubsection{Perturbation Generation and Candidates Selection}
Subsequently, we employ the approximated objective function to introduce minor perturbations to the input text, generating a set of adversarial candidates for both char and homo strategies that adhere to the specified imperceptibility constraints. We generate 100 adversarial candidates for both char and homo strategies.

Once the adversarial candidates are generated, we select the valid ones for the next iteration update, as illustrated in Fig.~\ref{proposeFig} (c). To accomplish this, we discard candidates that fail to meet the constraints specified in Equation~\ref{eq:problem} and then select the top three candidates based on their fitness scores for the next search iteration.

\section{TTSlow Objective Functions}
\label{TTSlow Objective Function}
Among TTS models, there are two prevalent mechanisms to determine the output speech length: length predictor for non-autoregressive models, and special end-of-speech token for autoregressive model.
To apply TTSlow on a wide range of TTS models, we propose distinct objective functions for these two length control mechanisms.
In this section, we introduce our proposed objective functions in detail.
Specifically, we delve into an autoregressive TTS model SpeechT5 \cite{ao2022speecht5}, and a widely adopted non-autoregressive TTS model VITS \cite{kim2021conditional} to explain the attack methodologies.

\subsection{Autoregressive TTS Model}
To gain a comprehensive understanding of how TTSlow attacks operate on an autoregressive TTS model, we illustrate the attack process targeting SpeechT5-tts in Fig.~\ref{speecht5} where Speecht5-tts processes both text and speaker embedding inputs. Its architecture comprises a text encoder pre-net, text encoder, speech decoder pre-net, speech decoder, and speech decoder post-net. The speech decoder pre-net processes the log Mel-filterbank input in an autoregressive manner \cite{ao2022speecht5}. Subsequently, we delve into the specifics of how TTSlow is achieved concerning speaker embedding and text input.

\subsubsection{TTSlow-spk on Autoregressive TTS Model}
When the given benign input is a speaker embedding, TTSlow-spk attacks the speaker embedding while keeping the text input unchanged. As depicted in Fig.~\ref{speecht5} (b), the text encoder pre-net initially transforms the unchanged text input into an embedding vector, which is then further converted to text representation via the text encoder \cite{ao2022speecht5}.
The X-vector \cite{snyder2018x} serves as the benign speaker embedding and is concatenated with the output of the speech-decoder pre-net, followed by a linear layer. Similarly, in Section~\ref{Gradient Optimization}, the adversarial speaker embedding is obtained by adding the benign speaker embedding and the perturbation, which is updated via gradient optimization.

The speech-decoder post-net comprises two modules. The first module aims to convert the decoder output to a scalar through a binary classifier for predicting the stop token \cite{ao2022speecht5} (scalar 0 for no-stop and 1 for stop), where TTSlow-spk is intended to encourage the prediction of non-stop tokens. The second module predicts the log Mel-filterbank by feeding the decoder output into a linear layer and five 1-D convolutional layers \cite{ao2022speecht5}.

To slow down the speech generation, we propose to make the victim TTS model generate longer speech by optimizing the decoder post-net output towards the conversion to no-stop tokens. Therefore, the TTSlow-spk objective in Equation~\ref{ttslow-spk loss} can be calculated as:
\begin{equation}
 \mathcal{L}_{\text{TTSlow-spk}} = \mathcal{L}_{\text{BCE}}(f(s+\delta),y), \quad  s.t. ||\delta|| \leq \epsilon,
\end{equation}
where s denotes input speaker embedding and $\mathcal{L}_\text{BCE}$ represents binary cross entropy loss between the generated scalar from the binary classifier and target scalar $y$ for non-stopping purposes.

\subsubsection{TTSlow-text on Autoregressive TTS Model}
When the given benign input is text, TTSlow-text attacks the text while keeping the speaker embedding input unchanged, as shown in Fig.~\ref{speecht5} (a). The benign text input undergoes iterative modifications through differentiable objective approximation in two strategies, char, and homo, as illustrated in Section~\ref{Differentiable Objective Approximation}.

For instance, in Fig.~\ref{speecht5} (a), the benign text input "I HAVE A PUPPY" is altered to "I HATE A PUPPY" by swapping the character "V" with "T" in the char strategy, and adjusted to "I HA\textbf{\textit{V}}E A PUPPY" by replacing "V" with "\textbf{\textit{V}}" in the homo strategy.  The objective of TTSlow-text is to encourage the Victim TTS model to generate non-stop speech, and the TTSlow-text loss in Equation~\ref{ttslow-text loss} can be computed as:
\begin{equation}
         \mathcal{L}_{\text{TTSlow-text}} = 
       \mathcal{L}_{\text{BCE}}(f(E(t+\delta)),y), 
       \quad  s.t. ||\delta|| \leq \epsilon,
\end{equation}
where t denotes input text and $E(\cdot)$ represents the text embedding vector.

In summary, our proposed TTSlow approach aims to discover the optimal perturbation $\delta$ that minimizes either the loss $\mathcal{L}_\text{TTSlow-text}$ or $\mathcal{L}_\text{TTSlow-spk}$ by encouraging the generation of stop token, thereby reducing inference efficiency.

\subsection{Non-autoregressive TTS Model}
\begin{figure}[t]
\centering
\includegraphics[width=88mm]{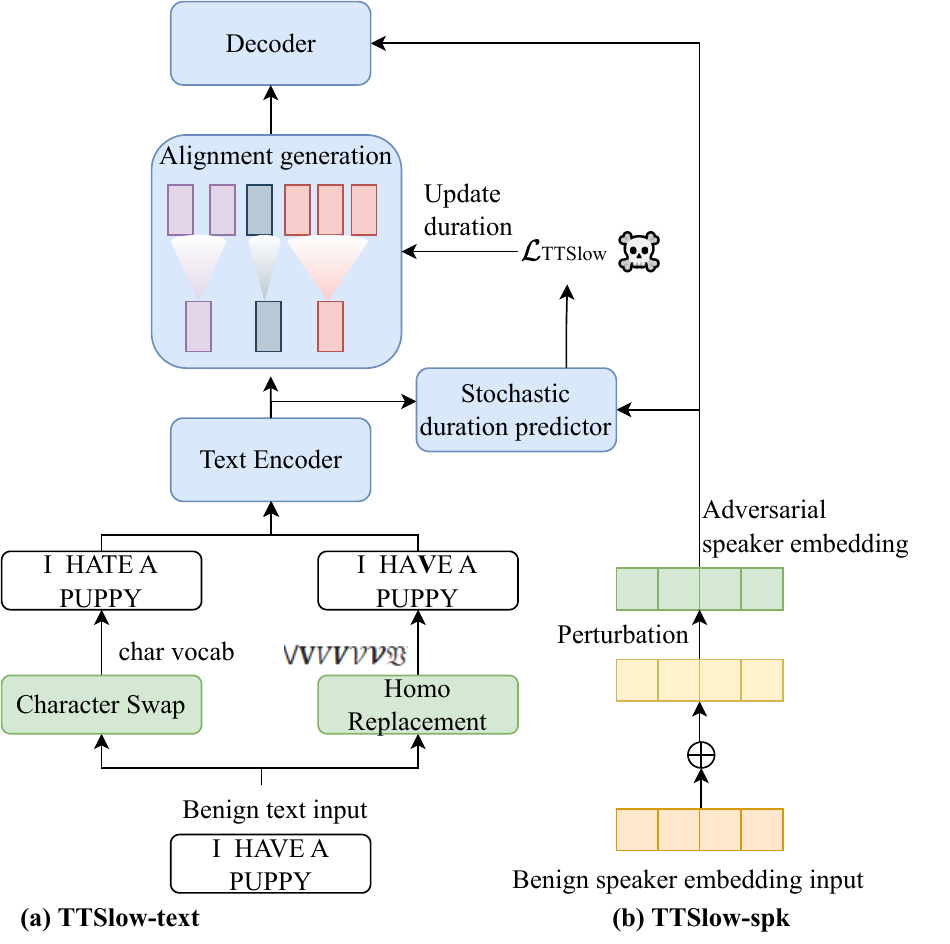}
\caption{Network Architecture of TTSlow with TTSlow-text and TTSlow-spk Attacks on the non-autoregressive VITS Model}
\label{VITS}
\end{figure}

Different from autoregressive model that generates speech in a sequential manner, non-autoregressive models are able to generate speech in parallel via duration predictor. We present the TTSlow attack process on the recent state-of-the-art non-autoregressive VITS model \cite{kim2021conditional,pratap2023scaling} in Fig.~\ref{VITS}. VITS is a parallel end-to-end architecture based on conditional variational auto-encoder (VAE), a stochastic duration predictor for alignment generation, Transformer-based encoder, and HiFi-GAN based decoder \cite{kim2021conditional}.

\subsubsection{TTSlow-spk on Non-autoregressive TTS Model}
When provided with a benign speaker embedding, TTSlow-spk attacks the speaker embedding of VITS while keeping the text input unchanged. The adversarial speaker embedding is also obtained by adding the benign speaker embedding and the perturbation, which is updated via gradient optimization as in Section~\ref{Gradient Optimization}. 
Subsequently, the updated adversarial speaker embedding is inputted into the decoder and stochastic duration predictor. The stochastic duration predictor plays a crucial role in estimating the distribution of text token durations and generating alignments for speech synthesis based on the inferred durations, as in Fig.~\ref{VITS} (b).

To this regard, we introduce a novel TTSlow-spk objective designed for non-autoregressive TTS models. The TTSlow-spk loss is formulated to maximize the cumulative output duration predicted by the stochastic duration predictor, thereby extending the length of the synthesized speech waveform. The TTSlow-spk objective, as defined in Equation~\ref{ttslow-spk loss}, can be computed as follows:
\begin{equation}
 \mathcal{L}_{\text{TTSlow-spk}} = -\sum{D(s+\delta)}, \quad  s.t. ||\delta|| \leq \epsilon,
\end{equation}
where s denotes input speaker embedding and $D$  signifies the output duration predicted by the stochastic duration predictor. Consequently, minimizing the TTSlow-spk loss is aimed at maximizing the output duration, ensuring continuous speech synthesis with non-stop purpose.

\subsubsection{TTSlow-text on Non-autoregressive TTS Model}
When provided with a benign text input, TTSlow-text attacks the text content while preserving the speaker embedding input, as depicted in Fig.~\ref{VITS} (a). The benign text input undergoes iterative modifications through differentiable objective approximation in two strategies, char, and homo, as illustrated in Section~\ref{Differentiable Objective Approximation}.  We introduce an innovative objective for TTSlow-text, which incentivizes the Victim TTS model to generate longer-duration speech by amplifying the duration outputs from the stochastic duration predictor. Thus, the TTSlow-text loss defined in Equation~\ref{ttslow-text loss} can be computed as:
\begin{equation}
         \mathcal{L}_{\text{TTSlow-text}} = 
        -\sum{D(E(t+\delta))}, 
       \quad  s.t. ||\delta|| \leq \epsilon,
\end{equation}
where t denotes input text and $E(\cdot)$ represents the text embedding vector.

In summary, our proposed TTSlow approach aims to discover the optimal perturbation $\delta$ by optimizing either the proposed objective $\mathcal{L}_\text{TTSlow-text}$ or $\mathcal{L}_\text{TTSlow-spk}$ via duration maximization to reduce the inference efficiency of the victim non-autoregressive TTS model.

\section{Experiments}
\label{Experiments}
In this section, we present datasets and experimental setup.
\subsection{Datasets}
For evaluation purposes, we utilize three widely recognized TTS datasets from Huggingface: the LibriSpeech dataset \cite{panayotov2015librispeech}, the LJ-Speech dataset \cite{ljspeech17}, and the English dialects database \cite{demirsahin2020open}, assessing them against multiple attack approaches. To manage computational demands, we evaluate the first 100 utterances from the LJ-Speech dataset\footnote{\url{https://huggingface.co/datasets/lj_speech}} and the first 100 utterances of the clean test subset from the LibriSpeech dataset\footnote{\url{https://huggingface.co/datasets/librispeech_asr}}. Additionally, we include the first 100 unique sentences from the Scottish female subset of the English dialects dataset \footnote{\url{https://huggingface.co/datasets/ylacombe/english_dialects}} for further evaluation. It's important to note that the LibriSpeech and LJ-Speech datasets are English datasets, while the English dialects dataset provides English accents, contributing to the diversity of our study.
\subsection{Experimental Setup}

\subsubsection{TTS Model Architecture and Attack Details}
For all attack approaches, we set the total number of iterations as 100 and beam size as 3. Learning rate $\alpha$ is set to 0.1 for TTSlow-spk, and we refer TTSlow-spk (l2) and TTSlow-spk (linf) with \( \ell_{2} \) norm and \( \ell_{inf} \) norm in Table~\ref{tableall}, respectively.
All models are implemented in Huggingface, where SpeechT5-tts is publicly available in the link~\footnote{\url{https://huggingface.co/microsoft/speecht5_tts}}. The encoder-decoder backbone in SpeechT5-tts contains twelve Transformer encoder blocks and six Transformer decoder blocks \cite{ao2022speecht5}. The text pre-net consists of a shared embedding layer, the speech-decoder pre-net and post-net use the same setting as in \cite{ao2022speecht5}. The HiFi-GAN vocoder \cite{kong2020hifi} is used to convert the log Mel-filterbank to the raw waveform.
Detailed parameters follow \cite{ao2022speecht5} and can be found in the source codes.
Target scalar $y$ is set to 0 to indicate no-stopping operation. 

Our experiments utilize the publicly available VITS-VCTK model~\footnote{\url{https://huggingface.co/kakao-enterprise/vits-vctk}} as the VITS victim model for speaker-oriented attacks (referenced in Table~\ref{tableall}). This model is trained using the VITS architecture on the VCTK dataset, which consists of approximately 44,000 short audio clips spoken by 109 native English speakers, totaling around 44 hours of audio \cite{kim2021conditional}. For speaker based attack approaches, we randomly select one speaker from 109 speakers to form the benign speaker embedding input. We also employ MMS-TTS model \cite{pratap2023scaling} as VITS victim model for text-oriented attack (also referenced in Table~\ref{tableall}), which is publicly accessible via the link~\footnote{\url{https://huggingface.co/facebook/mms-tts-eng}}. 
MMS-TTS builds upon the VITS architecture \cite{kim2021conditional} and extends its capabilities to support a triple-language setting across 1,107 languages. This advancement introduces greater challenges, given its increased power and awareness of large text corpora.

 \begin{table*}[t]
 \vspace{-0.3cm}
\caption{Comparison of the performance of different adversarial attack approaches on the SpeechT5 and VITS TTS models using the evaluation method, the number of frames under three datasets. Clean represents the original clean speech sample. The mean absolute value is computed by taking the average of generated speech samples for each dataset. The max absolute value is determined by selecting the highest value among \# frame values of the generated speech samples for each dataset.}
 \vspace{-0.1cm}
\begin{tabular}{lcll|rrrrr}
\label{table}
\\\toprule \midrule
\textbf{Evaluation}  & \textbf{Attack Model}  & \textbf{Datasets}    & \textbf{Attack Methods}   & \textbf{Mean Absolute} & \textbf{Max Absolute} & \textbf{Mean Incre} & \textbf{Max Incre}& \textbf{ASR (\%)}\\
\midrule
\# Frames &SpeechT5 & LJ speech & Clean   &    105,976    &    159,124& 0 & 0     & 0     \\
   \midrule
&SpeechT5   & LJ speech   & Text Baseline        &   138,409  &254,976  & 0.31& 0.60& 80 \\
&SpeechT5   & LJ speech   & Speaker Baseline &     90,179	  &180,736 &  -0.15   & 0.14 & 0 \\
   \midrule
&SpeechT5    & LJ speech   & TTSlow-text          &  189,926 & 360,960 &  0.79   & 1.27  &  98 \\
 &SpeechT5   & LJ speech   & TTSlow-spk (l2)        &  \textbf{424,156}	 &  \textbf{860,160}&   \textbf{3.00}   &  \textbf{ 4.41}& \textbf{98}\\
 &SpeechT5      & LJ speech   & TTSlow-spk (linf)   &	328,274  &860,160   & 2.10 & 4.41   & 97\\
   \midrule
   \# Frames   &SpeechT5  & Librispeech & Clean        &    107,290   &  373,040  & 0 & 0    & -    \\
   \midrule
&SpeechT5    & Librispeech & Text Baseline    &    141,932	      & 527,872   &    0.32 &   0.42 &  64 \\
&SpeechT5    & Librispeech & Speaker Baseline &     90,142	  &  381,440  &  -0.16   & 0.02&  5\\
   \midrule
&SpeechT5    & Librispeech & TTSlow-text      &      173,562	   &  533,504  &    0.62    & 0.43  &  86 \\
 &SpeechT5   & Librispeech & TTSlow-spk (l2)  &       \textbf{443,069}	  &   \textbf{1,817,600} &  \textbf{3.13}     & \textbf{3.87}  &  \textbf{97} \\
&SpeechT5    & Librispeech & TTSlow-spk (linf) &       394,281	  & 1,817,600   &   2.67  &   3.87  & 96\\\midrule
   \# Frames   &SpeechT5  & English Dialects & Clean        &   102,222	   &  212,992  & 0 & 0   & -     \\ \midrule
&SpeechT5   & English Dialects    & Text Baseline        &    117,365	  &  310,784  &     0.15   &     0.46 & 39 \\
&SpeechT5   & English Dialects    & Speaker Baseline &   73,344	  &  218,112 &    -0.28  &  0.02 & 2 \\
   \midrule
&SpeechT5    &English Dialects    & TTSlow-text          & 156,482  &335,360    &  0.53    &  0.57&  73  \\
 &SpeechT5   & English Dialects   & TTSlow-spk (l2)        &  \textbf{329,840}	 &  742,400&    \textbf{2.23}  & 2.49 & \textbf{96}\\
 &SpeechT5      & English Dialects   & TTSlow-spk (linf)   &    286,423	   &   \textbf{773,120} &   1.80 & \textbf{2.63} &95  \\
 \midrule \toprule
 \textbf{Evaluation}  & \textbf{Attack Model}  & \textbf{Datasets}    & \textbf{Attack Methods}   & \textbf{Mean Absolute} & \textbf{Max Absolute}              & \textbf{Mean Incre} & \textbf{Max Incre}& \textbf{ASR (\%)} \\\midrule
\# Frames &VITS  & LJ speech & Clean   &    105,976    &    159,124& 0 & 0   &-     \\
   \midrule
  &VITS   & LJ speech   & Text Baseline    &  104,709     & 166,400   & -0.01   &0.05 & 8\\
 &VITS   & LJ speech   & Speaker Baseline &     139,616    & 181,674 &    0.32  & 0.14 &54 \\
   \midrule
 &VITS    & LJ speech   & TTSlow-text      &     	156,121	 &284,160      &  0.47 &  0.79& 95\\
&VITS     & LJ speech   & TTSlow-spk (l2) & 174,295 & 315,648   & 0.64 & 0.98&\textbf{100}\\
&VITS     & LJ speech   & TTSlow-spk (linf)  & \textbf{301,839}  & \textbf{744,192} &   \textbf{1.85}   &   \textbf{3.68} &  \textbf{100}\\
\midrule
   \# Frames &VITS     & Librispeech & Clean   &     107,290   &  373,040  & 0 & 0   &   -  \\
   \midrule
&VITS     & Librispeech & Text Baseline    &    105,585	 & 320,256      & -0.12 &  -0.14&10\\
&VITS     & Librispeech & Speaker Baseline &     99,259   &  132,096& -0.07& -0.65 &49\\
   \midrule
 &VITS    & Librispeech & TTSlow-text      &       161,239&  	514,303 &  0.58 & 0.38& 91 \\
&VITS     & Librispeech & TTSlow-spk (l2)  &    166,019    &  537,600 &   0.55 &   0.44&88 \\
&VITS     & Librispeech & TTSlow-spk (linf) &  \textbf{264,020}    & \textbf{993,024}  &  \textbf{1.46}   &   \textbf{ 1.66} & \textbf{98}\\ \midrule 
   \# Frames   &VITS  & English Dialects & Clean        &     102,222& 	212,992    & 0 & 0     &-   \\ \midrule
&VITS   & English Dialects    & Text Baseline        &    90,071	&  169,216    &   -0.12     &  -0.21  & 9 \\
&VITS   & English Dialects    & Speaker Baseline &      88,399	& 107,562       &   -0.14   &  -0.49 & 26\\
   \midrule
&VITS    &English Dialects    & TTSlow-text          &  133,501	 &    295,168 &   0.31   &   0.39 & 59\\
 &VITS   & English Dialects   & TTSlow-spk (l2)        &  143,859	 &  323,072& 0.41     & 0.52 & 68\\
 &VITS      & English Dialects   & TTSlow-spk (linf)   &   \textbf{230,044}	    & \textbf{544,512}   &   \textbf{1.25} &  \textbf{1.56}&\textbf{92} \\ 
\bottomrule \bottomrule
\label{tableall}
\end{tabular}
\vspace{-0.2cm}
\end{table*}

\begin{table*}[t]
\caption{Comparison of adversarial text input samples and their decoded transcriptions from adversarially generated speech samples using Gemini, across different attacks: clean, text baseline, TTSlow-text, speaker baseline, TTSlow-spk (L2), and TTSlow-spk (Linf). Clean represents the original text of the clean speech sample.}
\vspace{-0.1cm}
\begin{tabularx}{\textwidth}{lXr}
\\ \toprule \toprule
\textbf{Attack} & \textbf{Adversarial text input samples}      & \textbf{\# Frames}                                                                                                             \\ \midrule
Clean          & especially as regards the lower-case letters; and type very similar was used during the next fifteen or twenty years not only by Schoeffer, &154,294 \\ \midrule
Text baseline           & esp\textbf{\textit{e}}cially a\textbf{\textit{s}} regard\textbf{\textit{s}} the lower-case \textbf{\textit{l}}etters; and type very similar was used dur\textbf{\textit{i}}ng the next fifteen or twenty y\textbf{\textit{e}}ars not only b\textbf{\textit{y}} Schoeffer, &\ 185,856\\ \midrule
TTSlow-text & espe\textbf{\textit{c}}ially as regards the lower-case letters; and type very similar w\textbf{\textit{a}}s used du\textbf{\textit{r}}ing the next fifteen or \textbf{\textit{t}}w\textbf{\textit{e}}nty years n\textbf{\textit{o}}t only by S\textbf{\textit{c}}hoeffer, & 299,520 \\ \midrule
Speaker baseline         & especially as regards the lower-case letters; and type very similar was used during the next fifteen or twenty years not only by Schoeffer, & 113,152\\ \midrule
TTSlow-spk (l2)         & especially as regards the lower-case letters; and type very similar was used during the next fifteen or twenty years not only by Schoeffer, & 581,632\\ \midrule
TTSlow-spk (linf)         & especially as regards the lower-case letters; and type very similar was used during the next fifteen or twenty years not only by Schoeffer, & 485,888\\ 
\toprule\midrule
\textbf{Attack} & \textbf{Decoded Transcription from Adversarially Generated Speech Samples via Gemini}      & \textbf{Score}   \\ \midrule
Clean          &  especially as regards the lower case letters And type very similar was used during the next 15 or 20 years Not only by Schaefer   & 9 \\ \midrule
Text baseline &   Especially regard the lower case A type very similar was used during in the next 15 or 20 years not only be show ever  &6\\ \midrule
TTSlow-text & S Ili as regards the lower case letters and type very similar double us used do in the next fifteen or twenty years in only by S Hoefer only by S Hoefer (noise) Ili as only by S Hoefer   &3 \\\midrule
Speaker baseline& especially as regards the lower case letters and type very similar was used during the next 15 or 20 years not only by Schaeffer  & 9\\\midrule
TTSlow-spk (l2) &  This bell (noise) Not only holy years not only by chauffeur holy years not only chauffeur chauffeur chauffeur chauffeur &2 \\\midrule
TTSlow-spk (linf) & especially as regards the lower case letters and type very similar was used during the next 15 or 20 years during the next 15 or 20 years not only by Schoeffer Then type very similar was used during the next 15 or 20 or 20 ors ors   & 5\\
\bottomrule \bottomrule
\end{tabularx}
\label{sample}
\end{table*}

\begin{figure*}[t]
\centering
\includegraphics[width=59mm]{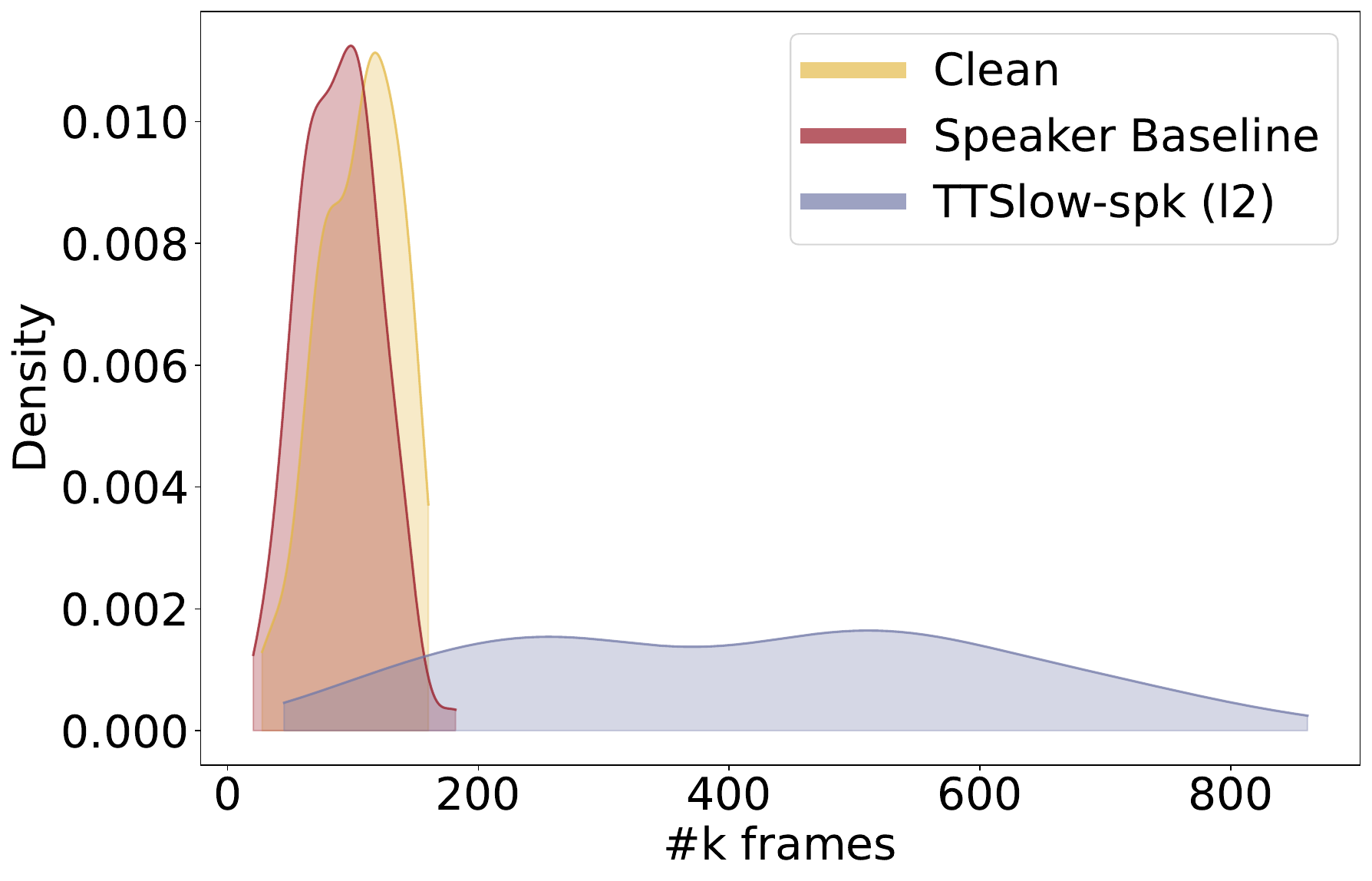}
\includegraphics[width=59mm]{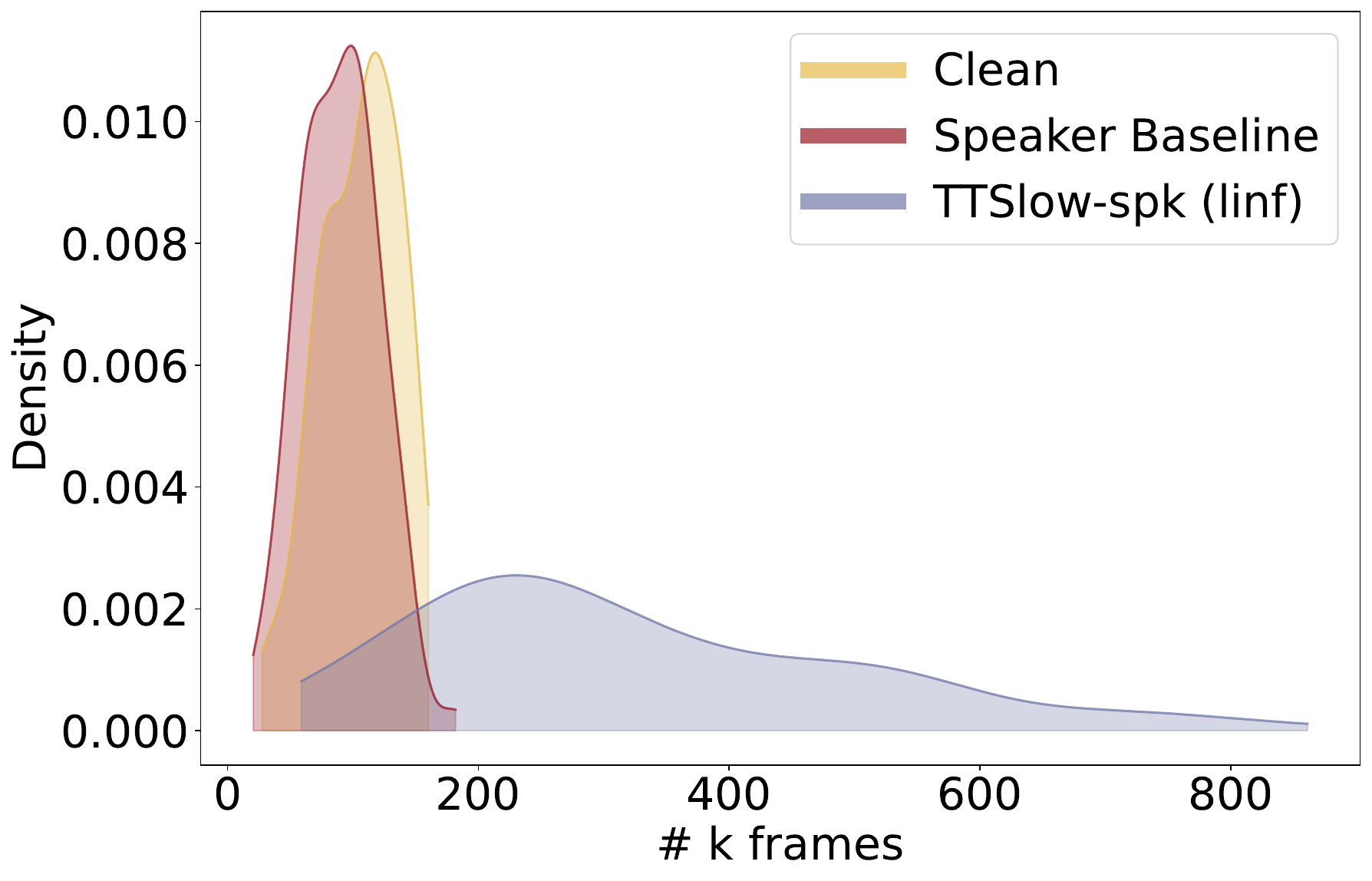}
\includegraphics[width=59mm]{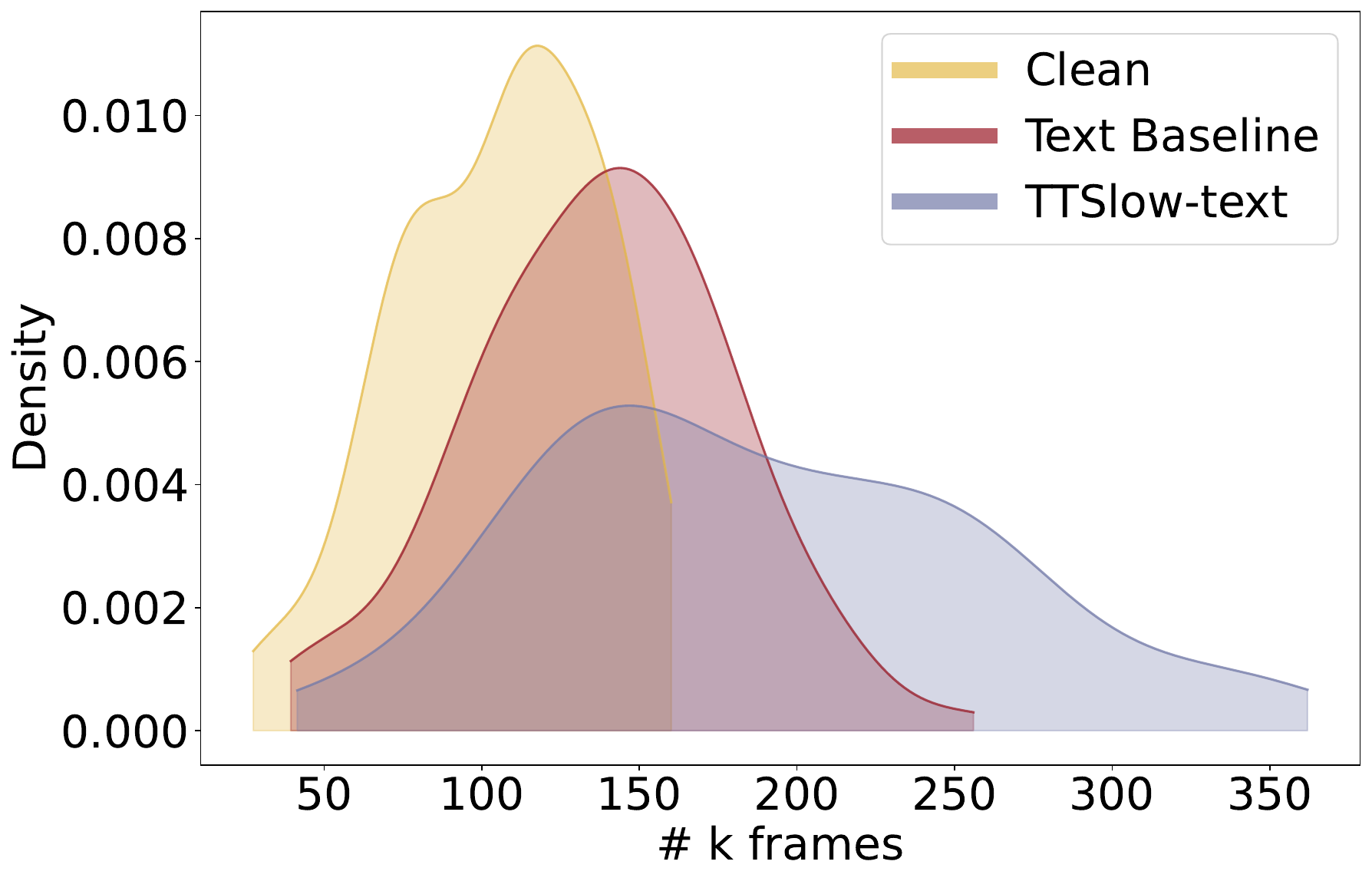}
\caption{Gaussian kernel density estimation (KDE) plots of speech length ($\#$ Frames) for clean data and the baseline approach with the proposed TTSlow-spk (l2), TTSlow-spk (linf), and TTSlow-text approaches on the LJ Speech dataset.}
\label{length}
\end{figure*}

\subsubsection{Baselines}
We present two baseline methods for each of our proposed attack techniques. In text-oriented attacks, the text baseline involves character swaps and homo replacements, without employing the iterative differentiable approximation optimization approach designed in our method. The number of characters replaced is determined by a fixed ratio of the input character length, set at 0.05. As for the speaker embedding attack baseline (speaker baseline), we utilize Gaussian perturbations without the proposed objective function and gradient optimization techniques.

\section{Results and Discussion}
\label{Results and Discussion}

\begin{figure*}[t]
\centering
\includegraphics[width=59mm]{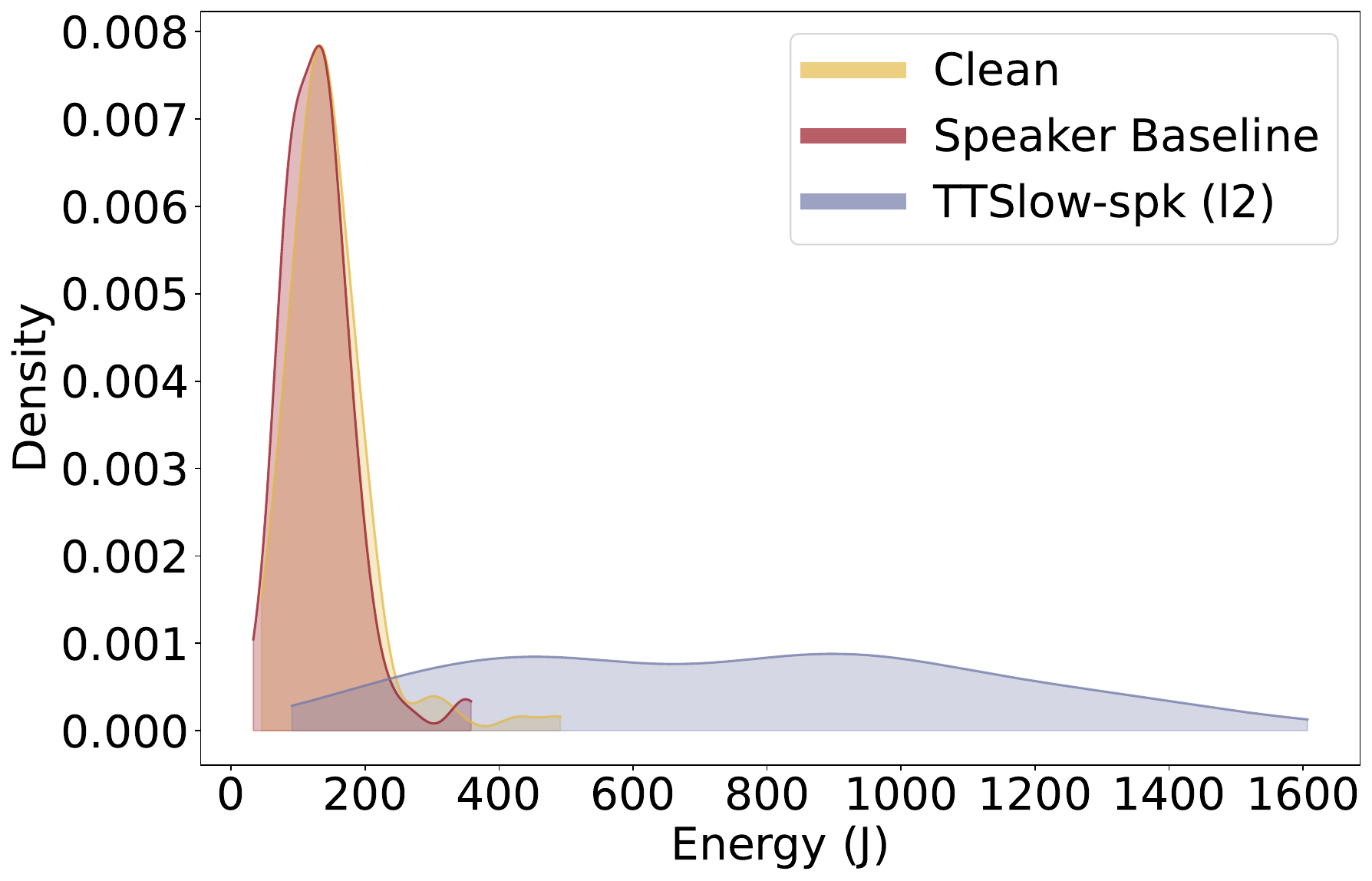}
\includegraphics[width=59mm]{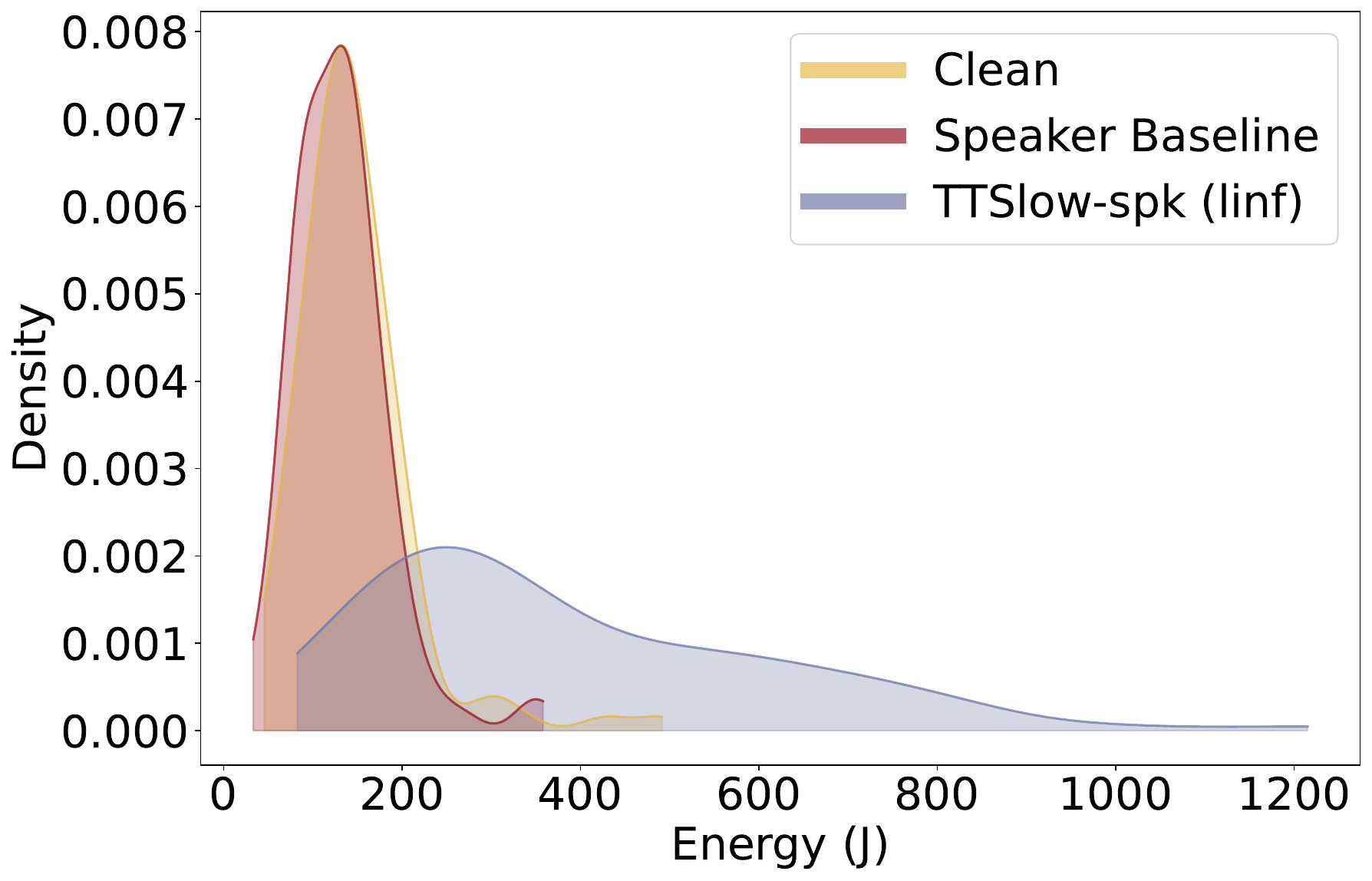}
\includegraphics[width=59mm]{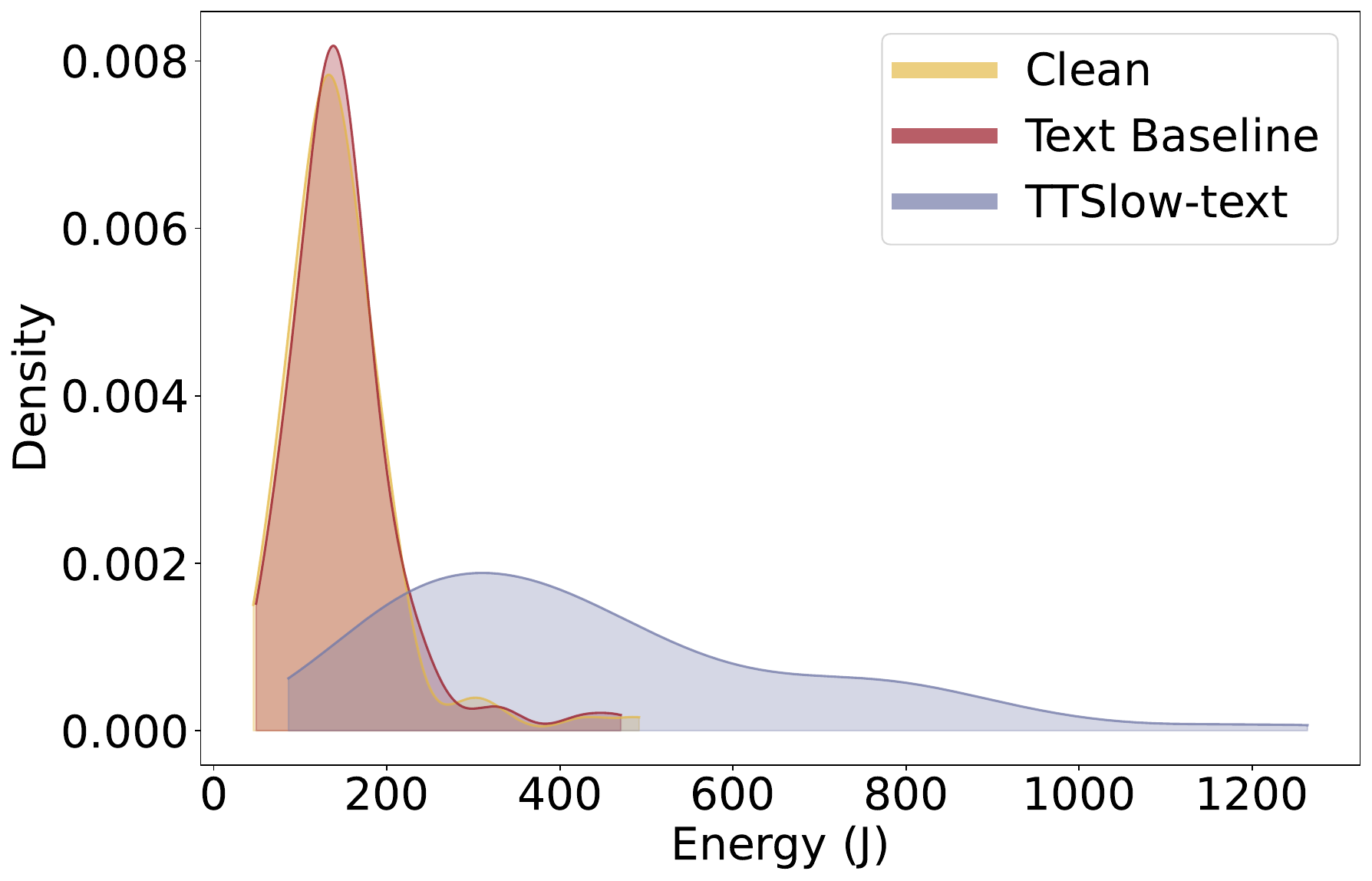}
\caption{Gaussian KDE plots of inference energy for clean data and the baseline approach with the proposed TTSlow-spk (l2), TTSlow-spk (linf), and TTSlow-text approaches on the LJ Speech dataset.}
\label{energy}
\end{figure*}

\begin{figure*}[t]
\centering
\includegraphics[width=59mm]{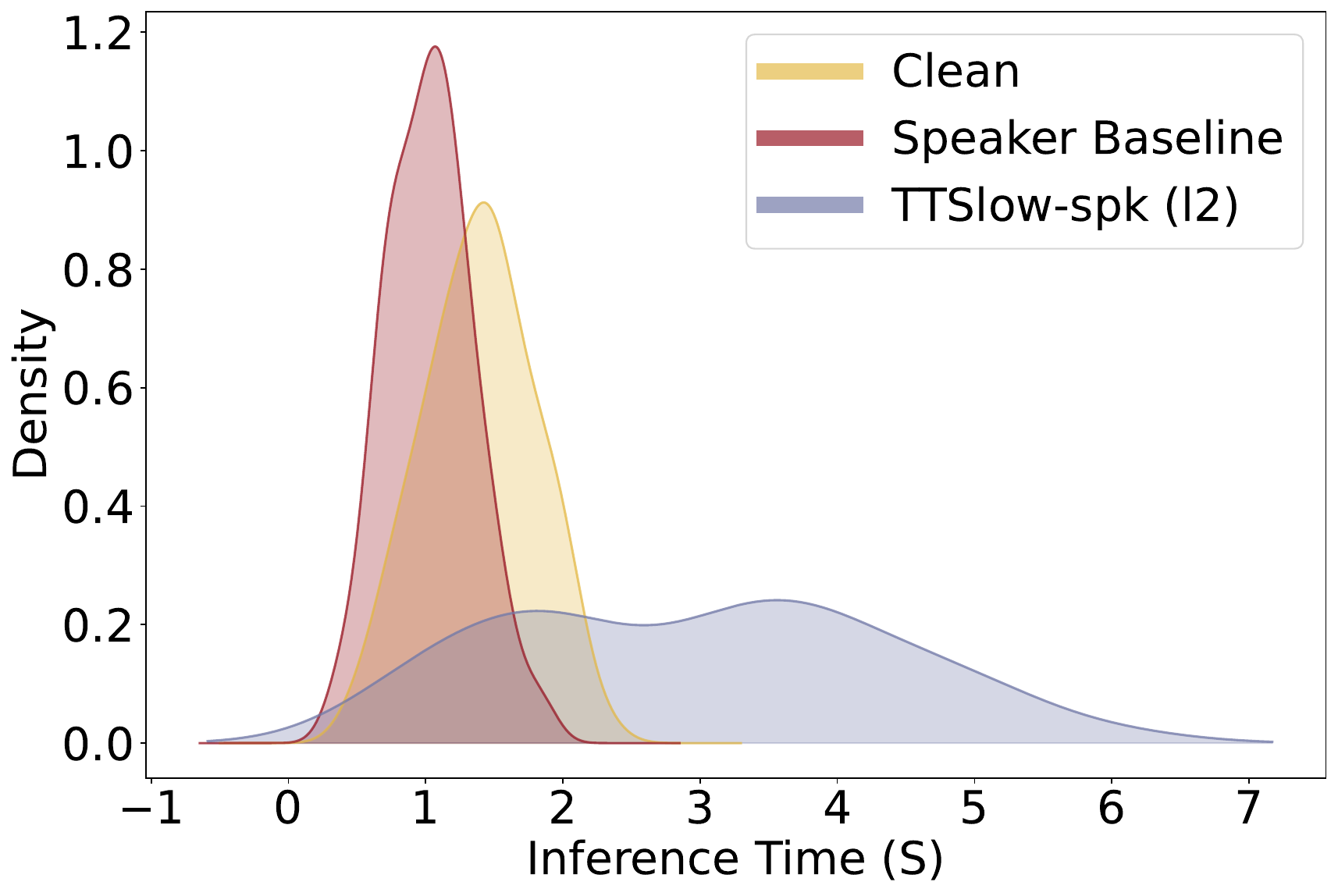}
\includegraphics[width=59mm]{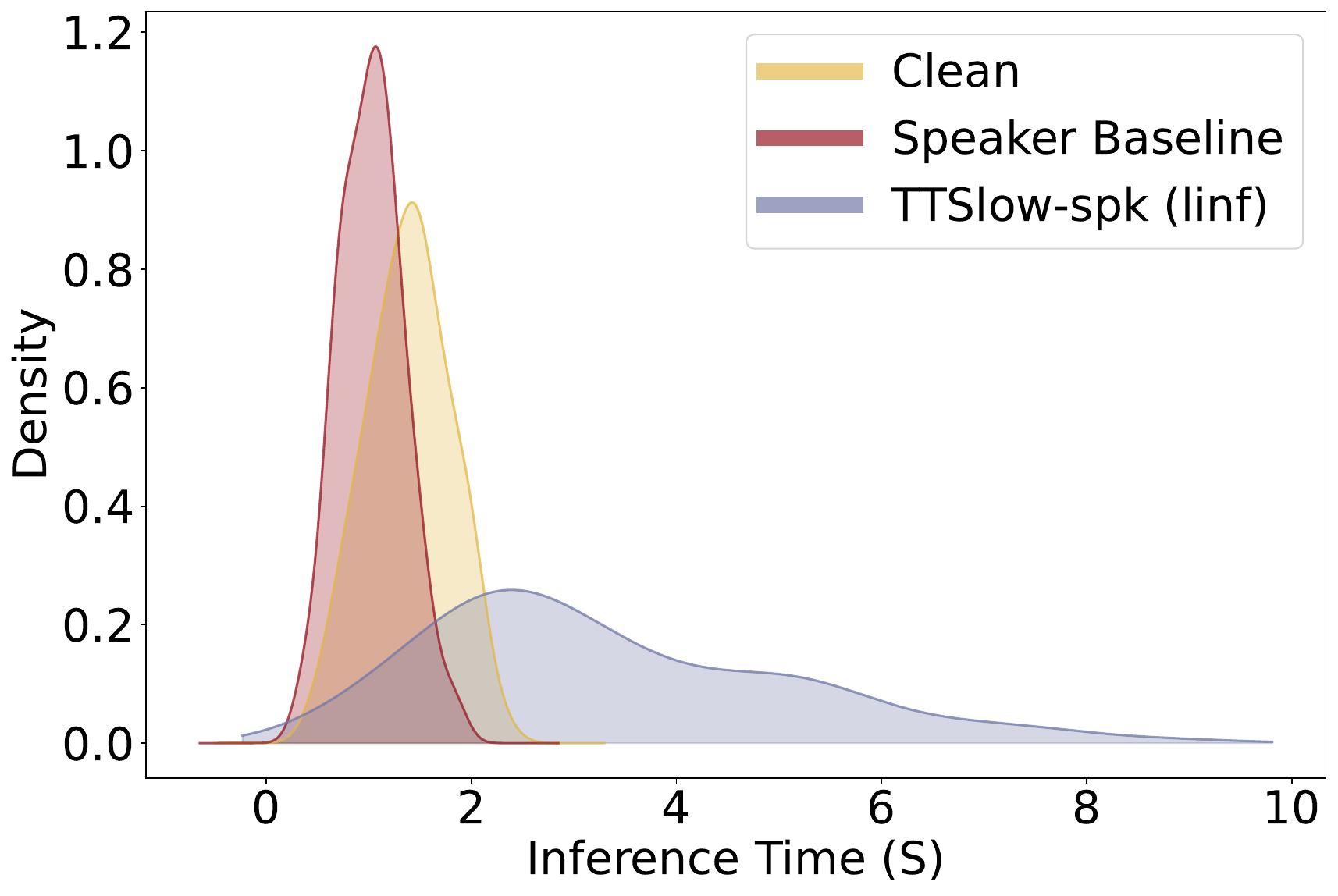}
\includegraphics[width=59mm]{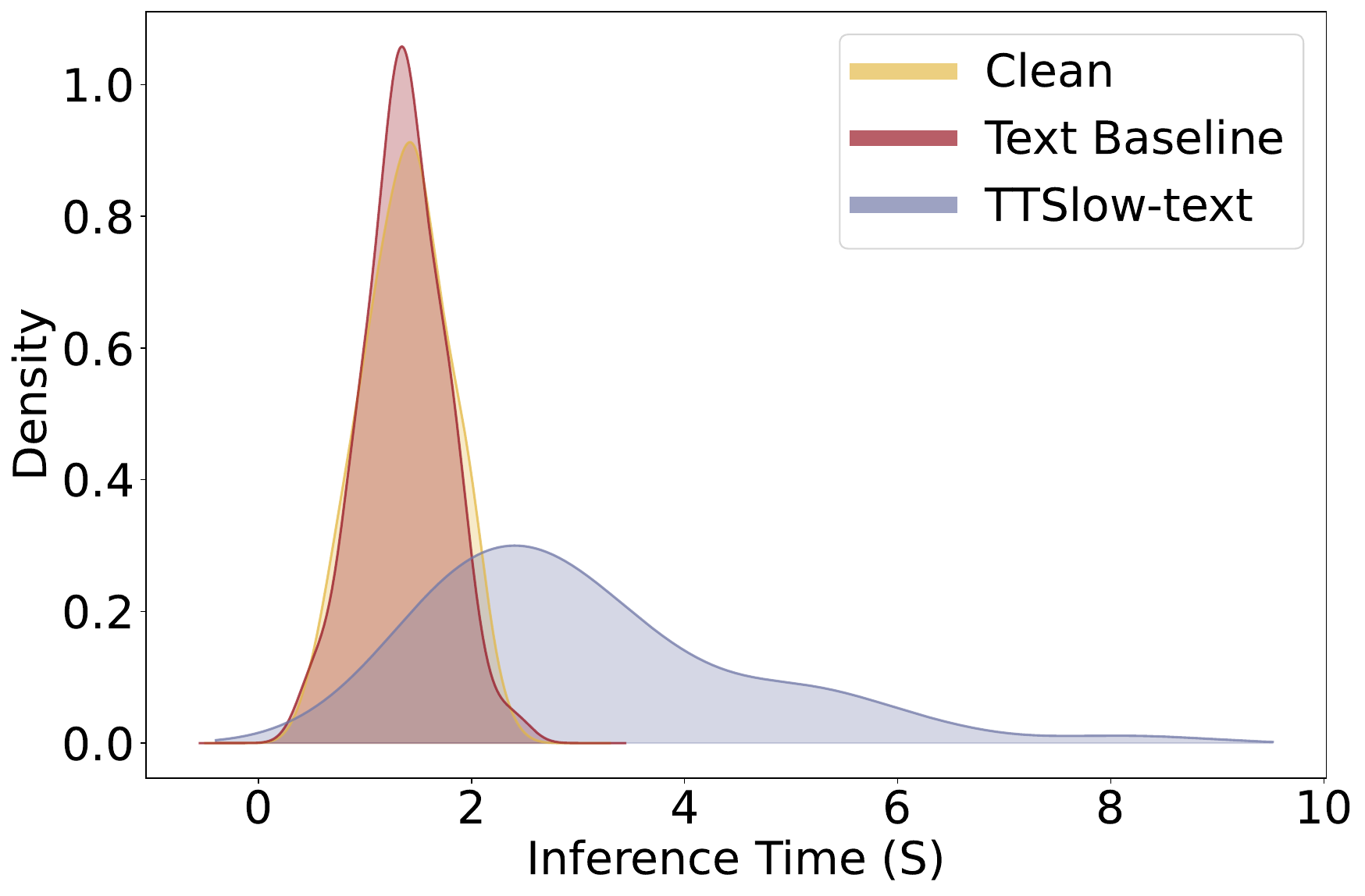}
\caption{Gaussian KDE plots of inference times for clean data and the baseline approach with the proposed TTSlow-spk (l2), TTSlow-spk (linf), and TTSlow-text approaches on the LJ Speech dataset.}
\label{time}
\end{figure*}

To evaluate the efficacy of our attack strategies, we quantify the output speech frames (\# Frames), with a higher count indicating better performance. Table \ref{tableall} presents the maximum and mean values of this metric for each dataset, along with the mean and max increment percentages to show improvements over original counterparts. Furthermore, we conduct a comprehensive analysis of inference efficiency in Fig.~\ref{time} and Fig.~\ref{energy}, including factors such as inference time and inference energy consumption, using visual methods.

We assess attack performance using the attack success rate (ASR), where a higher rate indicates better performance. ASR is calculated as the ratio of successfully attacked samples (where the adversarial speech length is 20\% longer than the original) to the total dataset size. Additionally, we provide adversarial samples for human imperceptibility analysis and include decoded transcriptions with their intelligibility evaluation by Gemini in Table~\ref{sample}.

\subsection{Vulnerability of the Victim TTS model}

Our objective is to investigate the vulnerability of a TTS model to adversarial attacks. As shown in Table \ref{tableall}, we observe that the TTS models are susceptible to all proposed adversarial attack techniques across three datasets. For instance, TTSlow-spk (l2) results in a relative 313 \% increase in the length of the speech sample compared to the original clean speech sample. This finding underscores the significance of evaluating the robustness of efficiency and designing defense systems for the victim TTS model.

\subsection{The Effectiveness of Adversarial Attacks}
We assess the efficacy of the proposed adversarial attacks by comparing them to their corresponding baselines for the number of frames. As shown in Table \ref{tableall}, both TTSlow-spk (linf) and TTSlow-spk (l2) significantly outperform the speaker baseline for all TTS models across all three datasets. Similarly, TTSlow-text consistently outperforms the text baseline for all TTS models across all three datasets. These findings demonstrate the effectiveness of the proposed attack approach. 

To provide a clearer visual representation of the length distribution in our generated speech samples, we present Gaussian kernel density estimation plots for the proposed TTSlow-spk (l2), TTSlow-spk (linf), and TTSlow-text, alongside their respective baselines and clean data in Fig.~\ref{length}. Observing the plot, we notice that the baseline densities occupy a larger area for longer speech frames than clean, while the proposed methods' densities exhibit a significantly larger footprint compared to both the baselines and the clean data. This observation suggests that the proposed methods effectively target longer speech outputs than the baseline attack methods and the original data.

\subsection{Attack Success Rate}

To comprehensively analyze the proportion of successful attacks on a per-dataset basis, we present the attack success rate (ASR) for all the models in Table~\ref{tableall}. We observed that our proposed TTSlow achieves higher ASR values compared to their baselines. Notably, TTSlow-spk even reaches 100\% ASR on LJ-Speech, indicating successful attacks on all samples. This underscores the effectiveness of our approach on TTS models across the three datasets, and identifies TTS models' weaknesses for security testing.

\subsection{Impact of Inference Efficiency}

To assess the inference efficiency of TTS models, we present Gaussian KDE plots of inference time and inference energy consumption of SpeechT5 model on LJ Speech data in Fig.~\ref{time} and Fig.~\ref{energy}, respectively. Note that the inference time and energy are all measured on the NVIDIA A40 GPU. We can see in Fig.~\ref{energy} that both the clean and baseline models consume a comparable amount of GPU energy, while the proposed TTSlow (shown in blue) significantly outperforms both the baselines and clean models. This demonstrates the effectiveness of our attack approaches in increasing inference energy and consequently reducing the overall inference efficiency of TTS models. 

Similarly, in Fig.~\ref{time}, we observe that the proposed TTSlow approach outperforms both the baselines and clean models significantly in terms of inference time. This suggests that TTSlow causes the TTS victim model to take longer for inference, leading to a decrease in inference efficiency. We observe from the figures that the proposed attacks not only extend the generated speech duration but also lead to higher computation, which indicates that the relevance of adversarial attacks on TTS systems lies in their potential to expose vulnerabilities and limitations within these systems. 

\subsection{Adversarial Samples}
To demonstrate the impact of the proposed adversarial perturbations, we present a case study of adversarial samples generated in Table~\ref{sample}. To enhance visibility, we utilize highlighted italic characters here to represent homoglyphs since the replacement of homoglyphs is challenging to discern with human eyes.
Due to space constraints, more adversarial samples generated using TTSlow can be accessed through the following link~\footnote{\url{ https://xiaoxue1117.github.io/TTSlow/}}. 

From the provided samples, we observe seven instances of homoglyph replacements for TTSlow-text ("c", "a", "r", "t", "e", "o" and "c"), resulting in approximately 61\% and 94\% longer speech samples compared to the text baseline and clean samples, respectively. We consider some of the attack examples are human imperceptible samples.
From the samples, we observe that the proposed attacks extend the duration of generated speech while maintaining nearly human imperceptibility.

\subsection{Evaluation with Gemini}
We conducted a case study to analyze error patterns in adversarial speech samples generated by the TTSlow attacker. This analysis included an understandability evaluation and transcription analysis of the synthesized speech samples using Gemini~\cite{reid2024gemini}, as shown in Table~\ref{sample}. The adversarial samples were transcribed by Gemini and assigned an understandability score from 0 to 10, with 10 indicating perfect understanding and 0 indicating no understanding. The transcriptions and their scores are presented in Table~\ref{sample}. We used Gemini 1.5 Pro, the latest large multimodal model for language, speech, and video~\cite{reid2024gemini}.

From Table~\ref{sample}, it is evident that the TTSlow approach yields lower understandability scores compared to clean and baseline samples. This outcome suggests that attacks successfully damage the content of the generated speech samples, making them challenging to understand. We also listened to the generated speech samples and observed error patterns, some of which can be found at the following link~\footnote{\url{ https://xiaoxue1117.github.io/TTSlow/}}.

Both the speech samples in the link and the decoded transcriptions in Table~\ref{sample} revealed several error patterns. These include word repetitions (e.g., "next 15 or 20 years" in TTSlow-spk (inf)), incorrect word generation (e.g., "S Hoefer" in TTSlow-text), and instances of long silence or noisy speech (e.g., noise in TTSlow-spk (l2) for example 3 in the above link). This error analysis highlights that TTSlow not only reduces TTS model inference efficiency but also achieves its accuracy-oriented attack goals by rendering the generated speech content incomprehensible through these error patterns.

\section{Conclusion}
\label{Conclusion}
In this work, we propose TTSlow, a novel adversarial attack approach that can decrease the efficiency of both autoregressive and non-autoregressive TTS models significantly. TTSlow enables diverse attacks on both speaker embeddings and text inputs, utilizing innovative objective functions and optimization techniques that have demonstrated effectiveness. By evaluating efficiency robustness in TTS models, our work contributes to advancing TTS research, bridging the gap between no attacks and multiple attack approaches while exploring inference efficiency robustness in TTS. Through extensive experiments on three publicly available datasets, we demonstrate that the proposed TTSlow approach outperforms baselines with improved performance. This study offers valuable insights for future research on efficiency robustness in TTS. In our future work, we plan to investigate accuracy-oriented attacks and advanced defense methods against adversarial attacks for TTS models.

\ifCLASSOPTIONcaptionsoff
  \newpage
\fi

\bibliographystyle{IEEEbib}
\bibliography{strings}
\newpage

\begin{IEEEbiography}[{\includegraphics[width=1in,height=1.25in,clip,keepaspectratio]{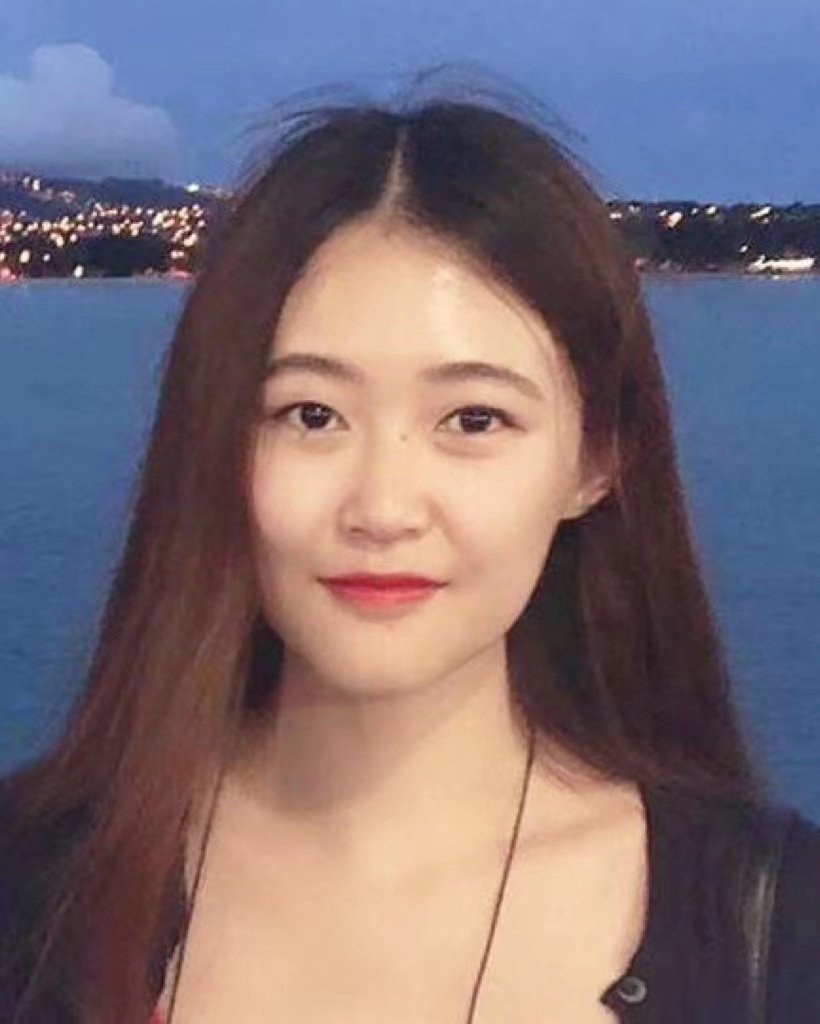}}]{Xiaoxue Gao}
 (Member, IEEE) received the
Ph.D. degree from the National University of Singapore (NUS), Singapore, in 2022, and B.Eng. degree in Electronic Information Science and Technology from Nanjing University, China, in 2017. She is currently a research scientist with Institute for Infocomm Research, Agency for Science, Technology, and Research (A*STAR). She has previously worked as a post-doctoral research fellow at NUS. Her research interests include speech recognition, self-supervised learning, speech synthesis, automatic lyrics transcription, audio security and multi-modal processing.
\end{IEEEbiography}
\vspace{-33pt}
\begin{IEEEbiography}
[{\includegraphics[width=1in,height=1.25in,clip,keepaspectratio]{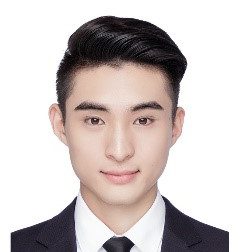}}]{Yiming Chen} received his E.Eng. degree in Computer Science and Technology from Southern University of Science and Technology, China, in 2020. He is currently a Ph.D. student of National University of Singapore, Singapore, since 2020. His research interests include natural language processing, model safety, and efficient machine learning.
\end{IEEEbiography}

\begin{IEEEbiography}
[{\includegraphics[width=1in,height=1.25in,clip,keepaspectratio]{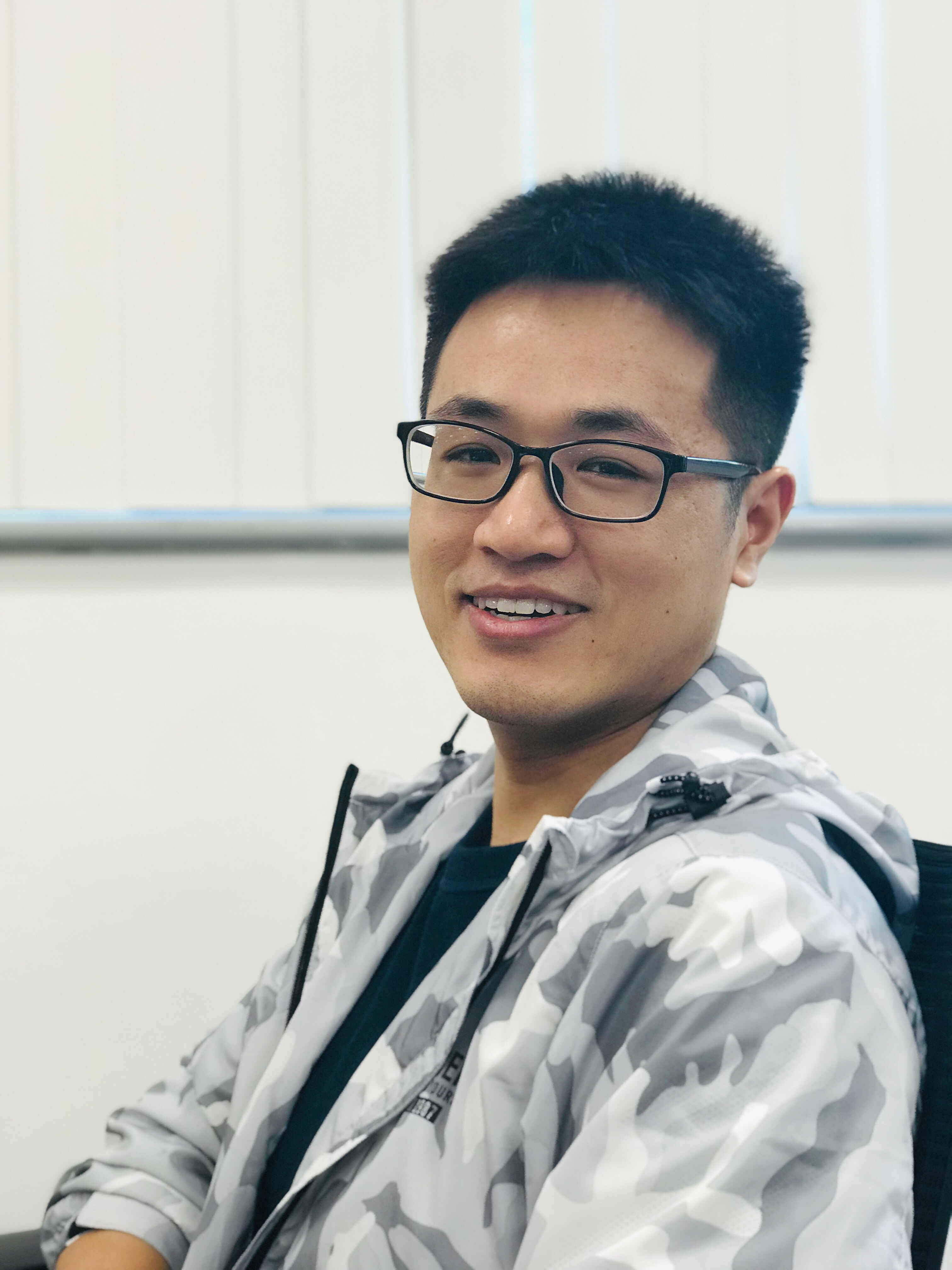}}]{Xianghu Yue} received the Ph.D. degree from the National University of Singapore (NUS), Singapore, in 2024, and B.Eng. degree in Automation from Beijing Institute of Technology, China, in 2016.
He is currently a research fellow at NUS. His research interests include speech recognition, self-supervised learning, and multi-modal processing.
\end{IEEEbiography}

\begin{IEEEbiography}
[{\includegraphics[width=1in,height=1.25in,clip,keepaspectratio]{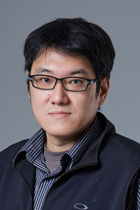}}]{Yu Tsao} (Senior Member, IEEE) received the B.S. and M.S. degrees in electrical engineering from National Taiwan University, Taipei, Taiwan, in 1999 and 2001, respectively, and the Ph.D. degree in electrical and computer engineering from the Georgia Institute of Technology, Atlanta, GA, USA, in 2008. From 2009 to 2011, he was a Researcher with the National Institute of Information and Communications Technology, Tokyo, Japan, where he engaged in research and product development in automatic speech recognition for multilingual speech-to-speech translation. He is currently a Research Fellow (Professor) and the Deputy Director with the Research Center for Information Technology Innovation, Academia Sinica, Taipei, Taiwan. He is also a Jointly Appointed Professor with the Department of Electrical Engineering, Chung Yuan Christian University, Taoyuan, Taiwan. His research interests include assistive oral communication technologies, audio coding, and bio-signal processing. He is currently an Associate Editor for the IEEE/ACM TRANSACTIONS ON AUDIO, SPEECH, AND LANGUAGE PROCESSING and IEEE SIGNAL PROCESSING LETTERS. He was the recipient of the Academia Sinica Career Development Award in 2017, national innovation awards in 2018– 2021, Future Tech Breakthrough Award 2019, Outstanding Elite Award, Chung Hwa Rotary Educational Foundation 2019–2020, and NSTC FutureTech Award 2022. He is the corresponding author of a paper that received the 2021 IEEE Signal Processing Society (SPS), Young Author, Best Paper Award.
\end{IEEEbiography}

\begin{IEEEbiography}
[{\includegraphics[width=1in,height=1.25in,clip,keepaspectratio]{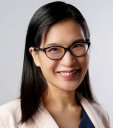}}]{Nancy F. Chen} (Senior Member, IEEE)  heads the multimodal generative AI group at A*STAR. She has served as the program chair of ICLR 2023, 2023 IEEE SPS Distinguished Lecturer, ISCA Board Member (2021-2025), in addition to being listed as 100 Women in Tech in Singapore 2021. Her honors include A*STAR Fellow (2023), the 2020 Procter \& Gamble (P\&G) Connect + Develop Open Innovation Award, the 2019 L’Oréal Singapore For Women in Science National Fellowship, Outstanding Mentor Award from the Ministry of Education in Singapore (2012), the Microsoft-sponsored IEEE Spoken Language Processing Grant (2011), and the NIH (National Institute of Health) Ruth L. Kirschstein National Research Award (2004-2008) and best paper awards from EMNLP, MICCAI, SIGDIAL, APSIPA, IEEE ICASSP. Speech evaluation technology developed by her team is deployed at the Ministry of Education in Singapore to support home-based learning during the COVID-19 pandemic. Nomopai is a spin-off company that uses technology from her lab to make customer agents more confident and empathetic. Prior to working at A*STAR, Dr. Chen worked at MIT Lincoln Laboratory on multilingual speech processing and obtained her PhD from MIT and Harvard University. 
\end{IEEEbiography}

\vfill

\end{document}